\documentclass[prl,aps,twocolumn,superscriptaddress,showpacs]{revtex4-1}

\usepackage{graphicx}% Include figure files
\usepackage{epstopdf}

\begin{document}

\title{Magnetic field splitting of the spin-resonance in CeCoIn$_{5}$}

\author{C. Stock}
\affiliation{NIST Center for Neutron Research, 100 Bureau Drive, Gaithersburg, Maryland 20899, USA}
\affiliation{Indiana University, 2401 Milo B. Sampson Lane, Bloomington, Indiana 47404, USA}
\author{C. Broholm} 
\affiliation{Institute for Quantum Matter and Department of Physics and Astronomy, Johns Hopkins University, Baltimore, Maryland USA 21218}
\affiliation{NIST Center for Neutron Research, 100 Bureau Drive, Gaithersburg, Maryland 20899, USA}
\author{Y. Zhao}
\affiliation{Department of Physics and Astronomy, Johns Hopkins University, Baltimore, Maryland USA 21218}
\author{F. Demmel}
\affiliation{ISIS Facility, Rutherford Appleton Labs, Chilton, Didcot, OX11 0QX}
\author{H.J. Kang}
\affiliation{NIST Center for Neutron Research, 100 Bureau Drive, Gaithersburg, Maryland 20899, USA}
\author{K. C. Rule}
\affiliation{Helmholtz Zentrum Berlin, D-14109, Berlin, Germany}
\author{C. Petrovic}
\affiliation{Condensed Matter Physics, Brookhaven National Laboratory, Upton, New York, USA 11973}

\date{\today}

\begin{abstract}

Neutron scattering in strong magnetic fields is used to show the spin-resonance in superconducting CeCoIn$_{5}$ (T$_{c}$=2.3 K) is a doublet.  The underdamped resonance ($\hbar \Gamma$=0.069 $\pm$ 0.019 meV) Zeeman splits into two modes at E$_{\pm}$=$\hbar \Omega_{0}\pm g\mu_{B} \mu_{0}H$ with $g$=0.96 $\pm$ 0.05.   A linear extrapolation of the lower peak reaches zero energy at 11.2 $\pm$ 0.5 T, near the critical field for the incommensurate ``Q-phase" (Ref. \onlinecite{Kenzelmann08:321}) indicating that the Q-phase is a bose condensate of spin excitons.

\end{abstract}

\maketitle

The presence of an underdamped resonance peak in the neutron scattering response has proven to be a strong indication of unconventional superconductivity where magnetism and electronic properties are strongly coupled.~\cite{Miyake86:34,Moriya00:49}  Spin resonances have been reported in a series of heavy fermion, cuprate and iron based superconductors and have been associated with the gap function undergoing a change in sign ($\Delta({\bf{q}}+{\bf{Q}_{0}})=-\Delta({\bf{q}})$). Therefore,  neutron scattering can be used to probe the electronic superconducting gap symmetry.   

It is to be expected that applied magnetic fields, which suppress the superconducting order parameter, should have a strong effect on the spin resonance.  Such effects have been difficult to pursue in the cuprates and iron based superconductors where chemical doping is required and resonance energies are high.~\cite{Dai00:406}  CeCoIn$_{5}$ is, however, particularly well suited owing to the stoichiometric nature of the compound and the accessible field and energy scales.

CeCoIn$_{5}$ displays an unconventional superconducting phase at ambient pressures and at temperatures below 2.3 K with a gap characterized by $d$-wave symmetry.~\cite{Petrovic01:13,Izawa01:87,Rourke05:94,Park05:72}   The structure is layered tetragonal with magnetic Ce$^{3+}$ ions in Ce-In(1) planes stacked along the c-axis and separated by a Co-In(2) network.~\cite{Kalyl89:1}   Despite the two-dimensional magnetic network, the Fermi surface is characterized by three dimensional sheets.~\cite{Hall01:64,Capan10:82,Settai01:13}   However, the superconductivity does reflect the underlying lamellar structure with a critical field of $\sim$12 T for fields applied with the $a-b$ plane and significantly lower $\sim$5 T for fields applied along $c$ at temperatures below 0.1 K.~\cite{Paglione03:91,Weickert06:74,Ikeda01:70,Tayama02:65}   

Neutron scattering [Ref. \onlinecite{Stock08:100}] shows the normal state has overdamped magnetic excitations peaked near {\bf{Q}}$_{0}$=(1/2,1/2,1/2) indicative of antiferromagnetic interactions between Ce$^{3+}$ ions both within the $a-b$ plane and along $c$.  The commensurate magnetic spin response differs from non-superconducting, though metallic, CeRhIn$_{5}$ which displays a magnetic Bragg peak at the incommensurate point {\bf{Q}}=(1/2,1/2,0.297) with a spiral magnetic structure.~\cite{Bao03:62}  On entering the superconducting phase in CeCoIn$_{5}$, an underdamped resonance peak at $\hbar \Omega_{0}$=0.60 meV  develops gathering spectral weight from low-energies.  These results indicated a strong coupling between f-electron $d$-wave superconductivity and magnetism.  A similar result and analysis has been applied to the heavy fermion superconductor CeCu$_{2}$Si$_{2}$ where a spin resonance has also been observed in the superconducting phase.~\cite{Stockert11:7}

While no magnetic Bragg peak was found at zero fields in CeCoIn$_{5}$, incommensurate order with {\bf{Q}}=(0.45,0.45,0.5) was observed for fields within the $a-b$ plane in a narrow field range below H$_{c2}$.~\cite{Kenzelmann10:104,Kenzelmann08:321,Blackburn10:105,Koutr10:104}  This was termed the ``Q-phase".  The new magnetic Bragg peak appears to be directly linked with superconductivity as it vanishes abruptly for magnetic fields above H$_{c2}$.  

The underlying structure of the resonance peak has been a matter of considerable theoretical interest.  One means of probing this is through high field spectroscopy which may lift any degeneracy of the resonance mode.  Here we demonstrate that the spin resonance peak in CeCoIn$_{5}$ is a doublet and the lower branch represents the soft mode of the ``Q-phase" order.

The results are based upon experiments performed on four cold neutron spectrometers.  The sample consisted of  $\sim$ 300 crystals aligned in $(HHL)$ reciprocal lattice plane as described previously.~\cite{Stock08:100}  High resolution measurements in a vertical magnetic field aligned along the [1$\overline{1}$0] direction were performed on the OSIRIS spectrometer (ISIS, UK) with a fixed E$_{f}$=1.84 meV.   By rotating the sample through $\sim$ 15 positions spaced 0.5$^{\circ}$ apart, a map in momentum and energy was constructed from which constant Q spectra near the commensurate (1/2,1/2,1/2) position were extracted.  Triple-axis measurements with vertical fields were also performed at SPINS and MACS (NIST, USA) with E$_{f}$=3.7 meV and 3.5 meV respectively.   Horizontal field measurements were taken at FLEX (Helmholtz Zentrum Berlin).   The field was within the $(HHL)$ plane rotated 30$^{\circ}$ from [001] to improve access for the incident and scattered beams.  For the horizontal field data discussed in this paper, we list the component of field projected along the $c$ axis.

\begin{figure}[t]
\includegraphics[width=8.5cm] {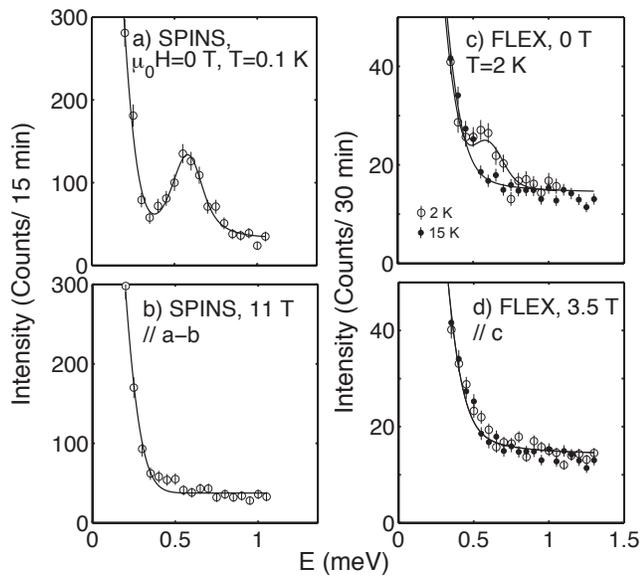}
\caption{\label{summary} Panel $a)$ and $b)$ illustrate the {\bf{Q}}=(1/2,1/2,1/2) resonance in zero field (T=0.1 K) and 11 T (near H$_{c2}$) along the [1$\overline{1}$0] direction.  Panels $c-d)$ are taken with the field applied along the $c$ axis.  The curves are fits to a simple harmonic oscillator and a background originating from incoherent elastic scattering.}
\end{figure}

The effect of magnetic fields, close to the upper critical field for superconductivity, on the spin resonance is summarized in Fig. \ref{summary}.  Panels $a)$ and $b)$ show results for the fields along [1$\overline{1}$0] where H$_{c2}$=12 T.  For 0T, we reproduce our previous results, while panel $b)$ shows that at 11 T a resonance is no longer observed.  Panel $c)$ demonstrates that we can still observe the resonance peak under the more constrained condition imposed by the horizontal field configuration on FLEX at 2 K. For modest fields along [001] near H$_{c2}$=5 T (panel $d)$, the resonance is suppressed, presumably replaced by the over damped fluctuations reported at similar fields by NMR.~\cite{Sakai11:107}  Therefore, the resonance peak is directly related to superconductivity and vanishes with the order parameter. 

\begin{figure}[t]
\includegraphics[width=7.7cm] {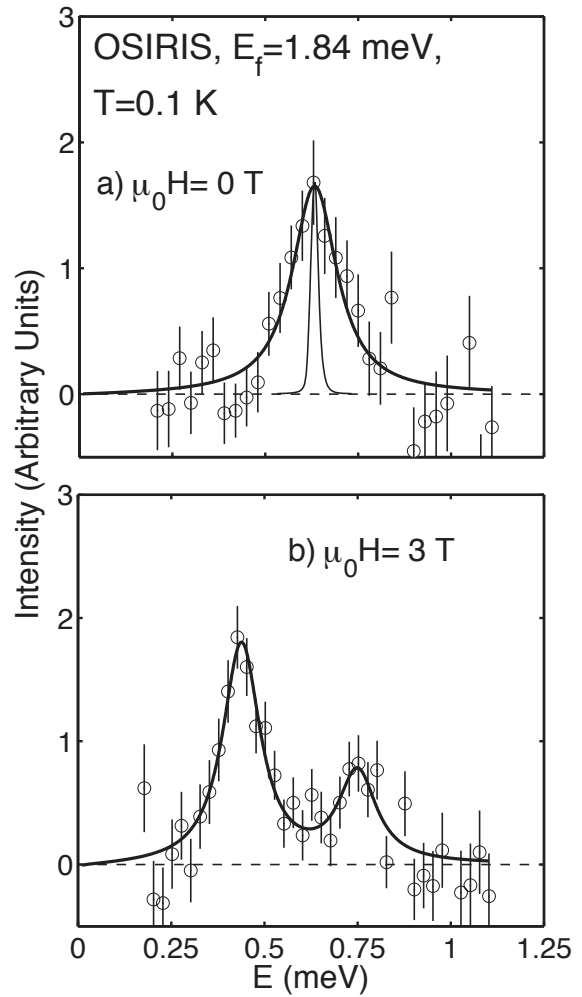}
\caption{\label{osiris} a) illustrates a high resolution scan through the spin resonance at zero applied field.  The solid curve centered at 0.6 meV illustrates the resolution function on OSIRIS with a full width of 0.025 meV.  b) demonstrates the splitting of the resonance into two peaks under an applied field of 3 T. A 10 K background has been subtracted from the scans.}
\end{figure}

Fig. \ref{osiris} illustrates the response of the spin resonance to intermediate fields in the superconducting phase well below H$_{c2}$.  The constant-Q scans are formulated by integrating around H=[0.45,0.55] and L=[0.45,0.55] on the OSIRIS indirect geometry spectrometer.  The scans were performed at T=0.1 K well below the transition to superconductivity (T$_{c}$=2.3 K) and the vertical field was applied along the [1$\overline{1}$0] axis with the sample aligned in the (HHL) scattering plane.  The resolution at the elastic line on OSIRIS is 0.025 meV (full-width at half maximum) and is illustrated by the solid curve in panel $a)$ centered at 0.63 meV.~\cite{Telling04:81}  A background derived from a similar scan at 10 K has been subtracted.  The solid lines are fits to harmonic oscillators convolved with the measured elastic resolution function.  While previous measurements on SPINS found the resonance peak width to be largely defined by the energy resolution of the spectrometer, panel $a)$ shows that the zero field resonance does have a finite lifetime with $\hbar\Gamma$= 0.069 $\pm$ 0.019 meV.   

Fig. \ref{osiris} $b)$ shows the same scan but in an applied vertical field of 3 T along the [1$\overline{1}$0] direction at 0.1 K.  The single peak observed in panel $a)$ at zero field is seen to be split into two peaks and this demonstrates that the resonance peak in CeCoIn$_{5}$ is a doublet.  The intensity ratio between the two peaks is 0.41 $\pm$ 0.11 at 3 T.  The width of the two peaks are equal, to within experimental error, and fitted to be 0.056 $\pm$ 0.008 meV.
\begin{figure}[t]
\includegraphics[width=6.7cm] {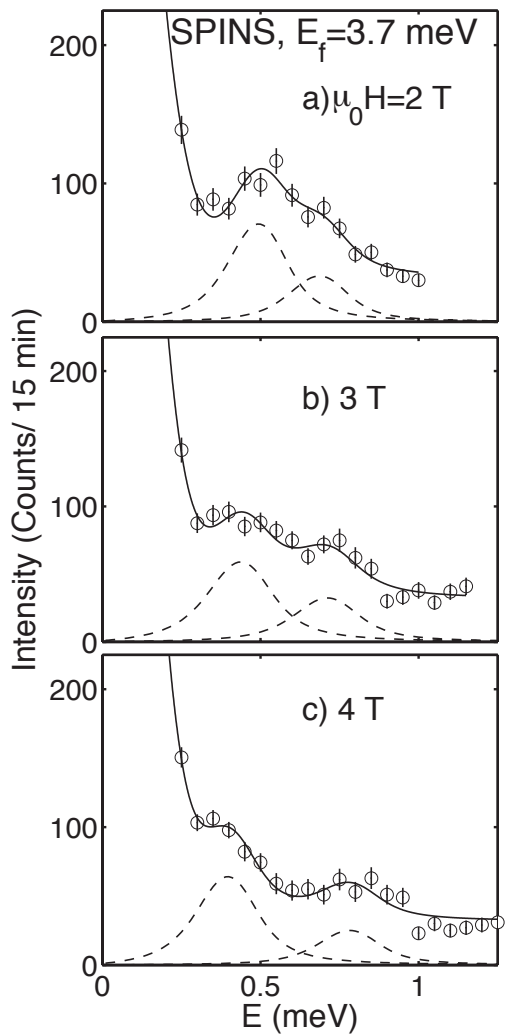}
\caption{\label{spins}The magnetic field dependence of the resonance peak at 0.1 K taken on SPINS.  The solid lines are fits to two harmonic oscillators and the dashed lines indicate the individual fits.  An overall background fixed to the 11 T scan has also been added.}
\end{figure}

For measurements over a broader range of fields, we use the coarser resolution and higher intensity of the cold neutron triple-axis spectrometer SPINS.  Fig. \ref{spins} illustrates the evolution of the resonance peak as a function of field at 2 T, 3 T and 4 T.  The solid line is a fit to a linear combination of two damped harmonic oscillators of equal width.  The data at 2 T (panel $a$) shows a broadening of the resonance which persists to 3 T and is consistent with Fig. \ref{osiris} with the larger resolution width of 0.15 meV.  At 4 T,  (panel $c)$ a distinct splitting can be resolved and two peaks are observed.  The intensity ratio at 4 T is 0.39 $\pm$ 0.1, consistent with the 3T OSIRIS data illustrated in Fig. \ref{osiris}.  These results are consistent with a previous cold triple-axis study (Ref. \onlinecite{Panarin09:78}) which tracked the softening of the lower peak with field, but did not observe the upper peak of the doublet shifted to higher energies.

\begin{figure}[t]
\includegraphics[width=7.5cm] {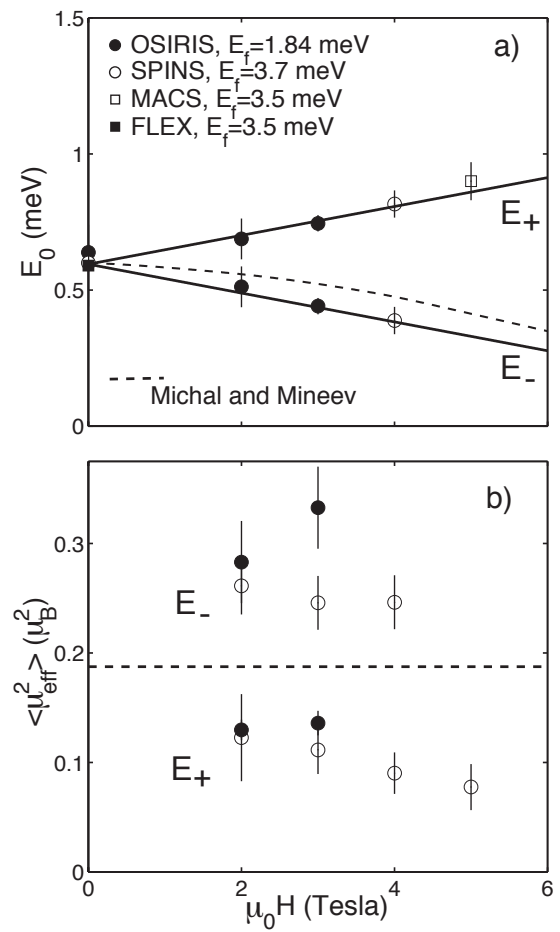}
\caption{\label{energy_position} $a)$ The peak position of the magnetic field splitting of the resonance peak as a function of applied field within the $a-b$ plane.  The solid lines are fits to $E_{\pm}=\hbar \Omega_{0} \pm g \mu_{B}\mu_{0}$H with $g=0.96\pm$0.05. The dashed line is the theory described in Ref. \onlinecite{Michal11:84}.  $b)$ illustrates the integrated intensity of the upper split peak in absolute units.  The dashed line is $1/2$ the integrated intensity of the zero field resonance peak.}
\end{figure}

We plot the peak positions (Fig. \ref{energy_position} $a)$) and intensities (Fig. \ref{energy_position} $b)$) as a function of magnetic field applied along the [1$\overline{1}$0] direction.    The solid lines are fits to $E_{\pm}=\hbar\Omega_{0}\pm g \mu_{B} \mu_{0} H$ as expected for a Zeeman split doublet.  The slope $g$=0.96 $\pm$ 0.05 maybe compared with the Lande factor of 0.83 for a free Ce$^{3+}$ ion and 0.81 calculated from a crystal field analysis with an in-plane magnetic field.~\cite{Christianson04:70}  This comparison illustrates that the spectral weight in the resonance originates from localized $4f$ electrons associated with the Ce$^{3+}$ ions.  The dashed line is the calculated energy position from Ref. \onlinecite{Michal11:84} normalizing the y-axis energy scale to match the zero field resonance (0.6 meV) energy and the horizontal axis to agree with the onset of magnetic order in the Q-phase (taken to be 10.6 T).  The intensity of the two peaks is displayed in Fig. \ref{energy_position} $b)$.  The $E_{+}$ peak shows a consistent trend to decreasing intensity at larger fields while the $E_{-}$ peak intensity is constant within error.  The dashed line is $1/2$ of the zero field resonance spectral weight.

Predictions for the field splitting of the resonance in the cuprates (in the context of Pr$_{0.88}$LaCe$_{0.12}$CuO$_{4-\delta}$) have suggested a splitting into three peaks reflecting an excitation from a singlet ground state to a triplet excited state.~\cite{Ismer07:99}  The central field independent peak is longitudinally polarized while the field dependent peaks are transverse.  In this theoretical study, the intensity of the two field dependent peaks was predicted to be equal.  However, in the close proximity of a particle-hole continuum the upper mode was predicted to weaken.  

The splitting of the CeCoIn$_{5}$ spin resonance is different from this scenario as only two field dependent peaks splitting with the Zeeman energy are observed.   Moreover, the upper peak progressively weakens with field (Fig. \ref{energy_position} $b$) and may indicate the resonance is located near a particle hole continuum causing a loss of integrated intensity of the upper excitation as this mode is driven into the continuum.  Such a scenario has been suggested in Ref. \onlinecite{Eremin08:101}, though other theories have been proposed (Ref. \onlinecite{Chubokov08:101}).  Missing spectral weight in a spin excitation also occurs in the cuprates at high energies and has been suggested to result from the close proximity of a continuum related to the pseudogap.~\cite{Stock07:75,Stock10:82}

The results for CeCoIn$_{5}$ differ from excitations observed in dimer quantum magnets (namely TlCuCl$_{3}$ and PHCC) where the ground state is a singlet and the first excited state is a triplet.~\cite{Ruegg03:423,Stone07:9}    The $\Delta S_{z}$=$\pm$1 modes have equal intensity and the $\Delta S_{z}$=0 mode (the central peak) is strongest.  In CeCoIn$_{5}$, only two field dependent peaks with differing intensity are observed.  Therefore, we cannot interpret the resonance as an excitation from a singlet ground state to an excited triplet state.

Any description of the resonance ($\hbar \Omega_{0}$=0.60 meV) must reconcile the experimental facts that the resonance is a doublet, the total spectral weight $\sim$ 0.37 $\mu_{B}^2$, and that a polarization analysis (based upon L scans using unpolarized neutrons) suggests the fluctuations are polarized along the $c$ axis corresponding to $J_{z}$ matrix elements.~\cite{weight}  One way of understanding this is to consider excitations from a superconducting $d$-wave condensate ($|\psi \rangle$) to an excited state that can be described as a condensate with a localized $4f$ spin ($| \psi, \pm\rangle$).  This exciton ~\cite{Michal11:84} state lies at an energy ($\hbar \Omega_{0}$) and is a doublet on account of the $4f$ crystal field environment.  Based on the zero field results, these two states are connected by $J_{z}$ but not by $J_{\pm}$ which presumably reflects a characteristic of the condensate.   In this picture the effect of the applied field would result in splitting of the doublet into two peaks E$_{\pm}$=$\hbar \Omega_{0}\pm g\mu_{B} \mu_{0}H$, with $g$ being the Lande factor for the localized $4f$ crystal field doublet.  This is consistent with the observation of two peaks and the experimental $g=0.96 \pm 0.05$.

Extrapolating the lower energy position in Fig. \ref{energy_position} $(a)$ to E=0 correspondingly suggests a quantum critical point at 11.2 $\pm$ 0.5 T, close to the field where the Q-phase is observed.~\cite{Kenzelmann08:321}  The spectral weight of the low-energy mode (Fig. \ref{energy_position} $(b)$) is similar to the 0.16 $\mu_{B}$ ordered moment reported for the Q-phase~\cite{Kenzelmann10:104} and the moments are aligned along $c$ as are the zero field resonant excitons.  Therefore, it appears the lower peak of the split doublet is the soft mode of the Q-phase which in turn can be interpreted as a bose condensate of $| \psi, \pm\rangle$ excitons.

We acknowledge funding from the STFC and the NSF through DMR-0116585 and DMR-0944772.  Work at IQM was supported by DoE, Office of Basic Energy Sciences, Division of Materials Sciences and Engineering under Award DE-FG02-08ER46544. Part of this work was carried out at the Brookhaven National Laboratory which is operated for the US Department of Energy by Brookhaven Science Associates (DE-Ac02-98CH10886).  We thank Z. Tesanovic and J. Murray for discussions and R. Down and E. Fitzgerald for cryogenic support.  We are grateful to Y. Qiu for altering DCS mslice to accommodate data taken on OSIRIS.

%\thebibliography{}
%\bibliographystyle{apsrev}
%\bibliography{CeCoIn5_field_bib}

\begin{thebibliography}{30}
\bibitem[{wei()}]{weight}
\bibinfo{note}{Based upon the crystal field analysis, we calculate the $J_{z}$
  matrix elements contribute 0.23 $\mu_{B}^{2}$ to $\langle \mu^{2}_{eff}
  \rangle$ and the transverse terms involving $J_{\pm}$, 1.9 $\mu_{B}^{2}$.}
\end{thebibliography}

%\end{thebibliography}

\end{document}